# Phase Diagram for Unzipping DNA Using a Bond-Force Criterion

Richard M. Neumann

*The Energy Institute, The Pennsylvania State University, University Park, PA 16802*

(January 29, 2002)

Using the criterion that the mechanical unzipping transition in a bound homopolymer is triggered when the average force exerted by the single unbound strands on the first base pair in the bound section exceeds the force binding the pair together, the temperature dependence of the critical unzipping force is obtained. In the resulting phase diagram, the critical force decreases monotonically with increasing temperature $T$; from a finite value at $T = 0$, to zero at the critical temperature.

Recently several authors[1-4] have examined the mechanical unzipping transition in the dsDNA homopolymer theoretically, using continuum models which analyze the polymer in the configuration shown in Fig. 1a. In this arrangement, the end of one of the two single strands of a dsDNA molecule is fixed while a constant force $\boldsymbol{F}$ acts on the end of the other strand. Establishing the criterion[2-4] for unzipping this system typically involves equating the free energy per monomer for the bound phase, $\lambda$, with that for the unbound phase, $\sim F^2/T$, where $T$ is the absolute temperature and $F = |\boldsymbol{F}|$. Here the Gaussian approximation is used to calculate the free energy for an unbound chain. Whereas these models do indeed predict an unzipping transition, calculating the temperature dependence of the critical force $F_c$ can be problematic. Depending on the spatial dimensionality of the model, $F_c$ can be zero at $T = 0$; it may increase with $T$ for all $T > 0$; or it may exhibit both characteristics, thereby yielding a phase diagram that is qualitatively different from that observed in experiments. Furthermore, when $\lambda$ is assumed to be $T$ dependent, the functional dependence is usually not specified, thereby preventing the construction of the phase diagram.

A phase diagram for dsDNA exhibiting a nonzero melting temperature, where $F_c = 0$, may be obtained in a straightforward manner using a three-dimensional model where the single strands are freely jointed chains, and the forces in the bond connecting the first base pair of the bound section are considered. Because the first base pair in the bound section of the DNA shown in Fig. 1a can be regarded as a Kuhn link of length $l$ in a chain consisting of sDNA links, the average force or tension $\langle f_l \rangle$ (derived from sources external to the base pair) present in this link can be determined simply by considering the link itself to be a dipole subject to the fixed-force couple $\boldsymbol{F}$, as depicted in Fig. 1b. We further assume that a threshold attractive force $f_a$ exists between the bases constituting this link, which, if exceeded in magnitude by an opposing "pulling" force, will result in the unzipping of the dsDNA. A precise description of this attractive interaction, which is temperature independent in our model, is not required; the interaction is regarded as an energy rather than a free energy.

The statistical-mechanical problem is thus reduced to one of calculating $\langle f_l \rangle$ as a function of $F$ and $T$, with the criterion $\langle f_l \rangle = f_a$ for the occurrence of the unzipping transition. The calculation[5] of $\langle f_l \rangle$ follows from the definition of the relevant partition function $Q$,

$$Q = \int d\mathbf{L} \, \exp(\beta \mathbf{F} \cdot \mathbf{L}) \delta(L - l), \qquad (1)$$

where $\beta = (k_B T)^{-1}$, and $d\mathbf{L} = L^2 \sin\theta \, d\theta \, d\phi \, dL$. $\theta$ and $\phi$ define the orientation of the "dipole" in a coordinate system whose $x$-axis is aligned with the direction of $\mathbf{F}$ and whose origin coincides with one end of the link, which is free to swivel while the force acts on the other end; see Fig. 1b. $Q \propto [l^2 \sinh(u)]/u$, where $u = \beta F l$ and $u_c = \beta F_c l$.

$$\langle f_l \rangle = k_B T (\partial \ln Q / \partial l) = 2k_B T/l + F\mathcal{L}(u), \qquad (2)$$

where $\mathcal{L}(u)$ is the Langevin function, $\coth(u) - 1/u$; $\mathcal{L}(u) = \langle \cos\theta \rangle$. Invoking the unzipping criterion, we are led to the following defining equation for $F_c$:

$$u_c \coth(u_c) = \beta l f_a - 1. \qquad (3)$$

Figure 2 provides the phase diagram in the force-temperature plane for the unzipping transition of dsDNA, as determined from Eq (3). Several conclusions from Eqs. (2) and (3) are worth emphasizing. At $T = 0$, $F_c = f_a$, as might be expected because in the absence of Brownian motion, no thermal force is available to "assist" in the mechanical unzipping; the external force $F$, alone, must overcome the base-pair binding energy. The critical (or melting) temperature $T_c$, determined from the condition $F_c = 0$, is equal to $lf_a/2k_B$. In the limit $T \to \infty$, $\langle f_l \rangle \to 2k_B T/l$; in other words, Eq. (2) states that with increasing temperature, the disruptive effect of the mechanical force on dsDNA decreases ($\langle \cos\theta \rangle$ decreases) whereas that of the Brownian force increases ($2k_B T/l$ increases).

The conventional continuum analysis of the unzipping transition yields[1-4]

$$F_c \propto [\lambda(T) T d]^{1/2} \qquad (4)$$

for temperatures in the moderate-to-high range; $d$ is the spatial dimensionality. For Eq. (4) to yield a $T_c$-displaying phase diagram, i.e., $F_c = 0$ at $T = T_c > 0$, it is necessary for $\lambda(T)$ to be a free energy with the property that $\lambda(T_c) = 0$. $\lambda$'s with this property are not readily derivable from the typical potential-energy functions used to describe the base-pair interaction in continuum models. On the other hand, with the approach advocated here, the destabilizing effect of the heat bath on dsDNA arises naturally in Eq. 2, as expressed in the $T/l$ term, which, in turn, leads to the phase diagram shown in Fig. 2. Finally, it is precisely in this term that the present work differs from previous apporaches. Omission of $2k_B T/l$ from Eq. 2 results in Eq. 4 (in the high-temperature, low-force range), with the understandng that $\lambda = lf_a$.

### Acknowledgment

The author has benefited from conversations with S. M. Bhattacharjee.

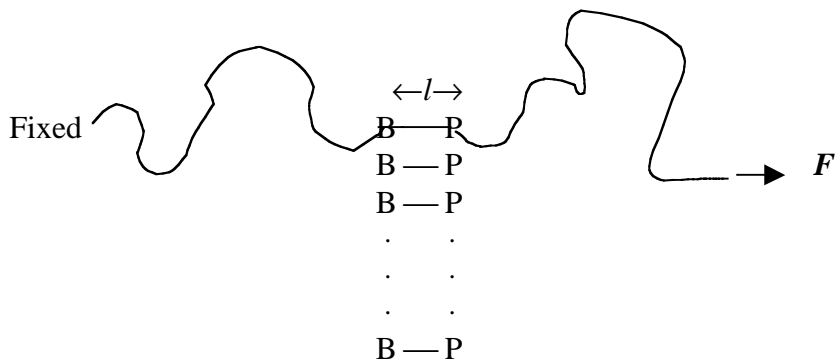

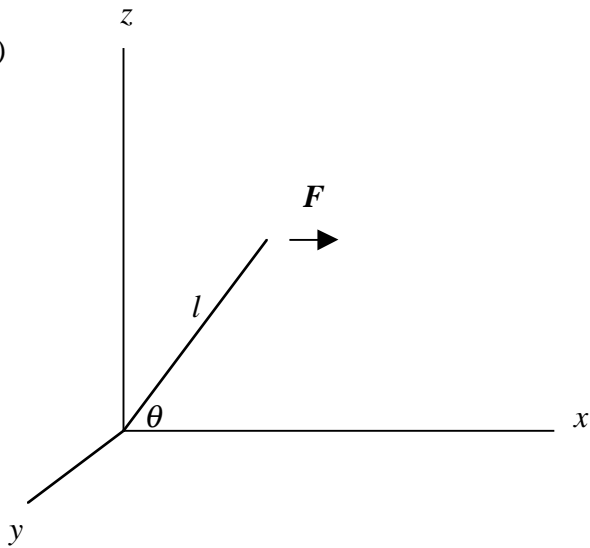

**Figure 1.** (a) The first bound base pair of length $l$ is subjected to a constant force $F$ by means of the unbound strands in a section of dsDNA; B—P denotes a base pair. (b) The link-dipole is shown in a coordinate sytem where the force $F$ is parallel to the $x$-axis; $\theta$ is the angle between the dipole and the $x$-axis.

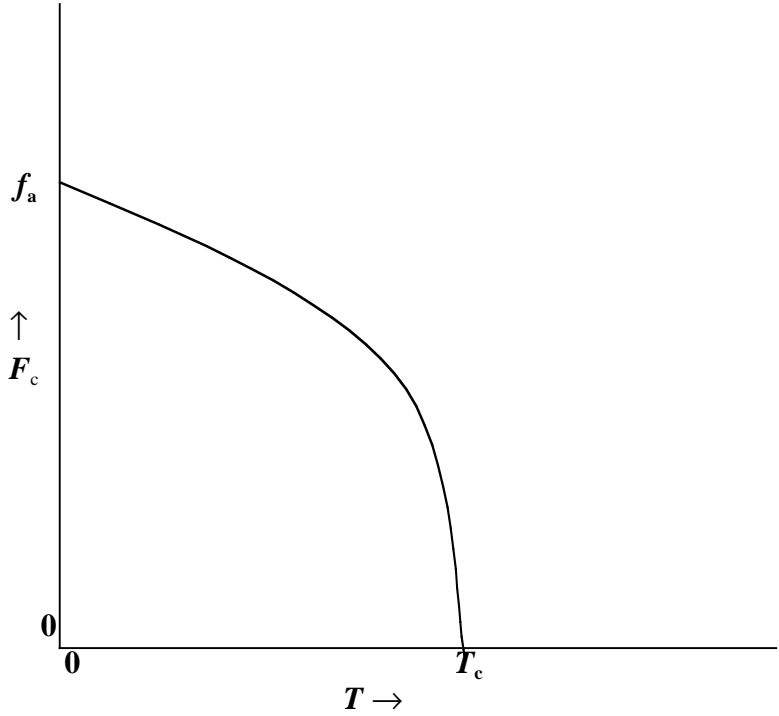

**Figure 2**. The critical force $F_c$ is shown as a function of the temperature $T$, based on Eq. (3), for the mechanical unzipping transition in a dsDNA homopolymer.